\documentclass[epj]{svjour}
\usepackage{graphicx}
\usepackage{amsmath}
\begin{document}
\title{Quantitative expression of the spin gap via bosonization for a
  dimerized spin-1/2 chain} 
\author{E. Orignac}
\institute{Laboratoire de Physique Th\'eorique de l'\'Ecole Normale
  Sup\'erieure CNRS--UMR8549  24, Rue Lhomond F-75231 Paris Cedex 05
  France}
\date{\today}
\abstract{
Using results on the mass gap in the sine-Gordon model combined with
the exact amplitudes in the bosonized representation of the Heisenberg
spin-1/2 chain and one-loop renormalization group, we derive a
\emph{quantitative} expression for the gap in a dimerized spin-1/2
chain. This expression is shown to be in good agreement with recent
numerical estimates when a  marginally
irrelevant perturbation is taken into account.} 
\maketitle

Low dimensional antiferromagnets have been the subject of intense
scrutiny both theoretical and experimental for the last twenty years. 
The simplest model, the spin-1/2 Heisenberg antiferromagnet, is
integrable\cite{bethe_xxx} and  can be
mapped onto a continuum field
theory\cite{luther_spin1/2,haldane_xxzchain,nijs_equivalence} which allows the full
determination of its zero temperature critical behavior. The presence
of  a marginally irrelevant operator in the continuum theory 
induces logarithmic corrections to the critical
scaling\cite{affleck_log_corr}. The corrections to scaling of the 
 correlation
 functions\cite{affleck_log_corr,giamarchi_logs,barzykin99_logs}, 
NMR relaxation rates \cite{barzykin01_nmr_logs,brunel99_edges_logs}, 
and susceptibilities\cite{eggert_susceptibility,starykh97_logs} 
in this model  have
been investigated in details. Further, when the Heisenberg spin-1/2
chain  model is  perturbed by a relevant operator such as an
alternation of the exchange coupling, the  marginal
operator gives rise to a  
logarithmic correction to the power law dependence
of the gap on the perturbation\cite{affleck_log_corr}. 
Such logarithmic corrections to scaling in the gap 
 in the context of
two dimensional statistical
mechanics of the four state Potts model whose transfer matrix is
related to the Hamiltonian of the alternating Heisenberg
chain\cite{nauenberg80_potts,kadanoff_ashkin,black_equ,kohmoto81_ashkin}. 
In \cite{black_equ} in particular, 
it was shown that the  dependence of the gap
$\Delta$ on the 
dimerization $\delta$, was changed from the form $\Delta \sim
\delta^{2/3}$ \cite{cross_spinpeierls,haldane_dimerized} to a the form $\Delta \sim
\delta^{2/3}/|\ln \delta|^{1/2}$.   
Such logarithmic behavior was confirmed by
numerical calculations in
\cite{spronken86_dimerized,kung86_dimerized,barnes_dimerized}.
Alternatively, the dependence of the gap on the dimerization can be
described by an effective power law form with an exponent that
deviates from $2/3$\cite{uhrig99_spinons,singh99_dimerized}. For a
dimerization not too small, 
it is found that the resulting effective exponent is
close to $2/3$ \cite{uhrig99_spinons}. Further, by
considering a Heisenberg chain with an additional next-nearest
neighbor coupling finely tuned to cancel the marginal operator, a pure 
power law with exponent $2/3$ can be obtained obtained for the gap
\cite{chitra95_dmrg_frustrated}. Recently, 
the logarithmic corrections were investigated in greater details using the
DMRG\cite{papenbrock_dimerized}. The data for the gap could be fitted
to the form:
\begin{eqnarray}
  \label{eq:gap_papenbrock}
  \Delta=\alpha_{gap}^{1/2} \frac{\delta^{2/3}}{(\ln \delta_0/\delta)^{1/2}}, 
\end{eqnarray}
\noindent with $\alpha_{gap}=19.4$ and $\delta_0=115$ 
or alternatively  by the power law form $\Delta=1.94\delta^{0.73}$.   
A difficulty that arises when comparing the predictions of the
Renormalization Group
approach\cite{affleck_log_corr,nauenberg80_potts,kadanoff_ashkin,black_equ,kohmoto81_ashkin}
 with the numerical results is that the former approach can only
 predict the exponents, and not the non-universal prefactors. 
However, exact results for the sine-Gordon model combined with recent
progress\cite{affleck_ampl_xxz,lukyanov_ampl_xxz}  on the bosonization
treatment of the Heisenberg spin 1/2
chain using integrability make it possible to overcome these two
difficulties and obtain the prefactor in the expression of the 
gap at least in the absence of logarithmic corrections.  
Assuming that the gap varies continuously as the marginally irrelevant
is turned on, it is then possible
to obtain an expression of the gap as a function of the marginally
irrelevant operator, with no further unknowns. Fitting the data of
\cite{papenbrock_dimerized} then allows the determination of the order
of magnitude of the marginally irrelevant interaction. The obtained
value can then be checked against the one obtained in
\cite{affleck_log_corr}. A similar approach  has been used previously
in \cite{hikihara03_amplitude_xxz} to estimate the gaps induced by a
staggered field in an anisotropic spin 1/2 chain.   
 
  The Hamiltonian of the dimerized spin-1/2 chain reads:
\begin{eqnarray}
  \label{eq:dimerized-spinchain}
  H=J \sum_n (1+(-)^n\delta) \mathbf{S}_n\cdot\mathbf{S}_{n+1} 
\end{eqnarray}
For $\delta=0$, this Hamiltonian reduces to the one of the uniform
antiferromagnetic Heisenberg chain ($J>0$) the low energy properties
of which are
described by the following continuum
Hamiltonian\cite{nijs_equivalence,black_equ}:
\begin{eqnarray}
  \label{eq:bosonized-ham}
  H=\int \frac{dx}{2\pi} \left[ uK (\pi \Pi)^2 + \frac u K (\partial_x
    \phi)^2\right] -\frac{2g_2}{(2\pi a)^2} \int dx \cos \sqrt{8} \phi,  
\end{eqnarray}
\noindent where $[\phi(x),\Pi(x')]=i\delta(x-x')$, $u=\frac \pi 2 J a$
and
$K=1-\frac{g_2}{2\pi u}$. The latter conditions ensures $SU(2)$
symmetry, and for $g_2<0$ the operators $(\pi \Pi)^2-(\partial_x
\phi)^2$ and $\cos \sqrt{8} \phi$ are both marginally irrelevant
resulting a gapless fixed point. The exact bare value of $g_2$ has
been estimated in \cite{affleck_log_corr}.    

\begin{figure}[htbp]
  \begin{center}
    \includegraphics[width=9cm]{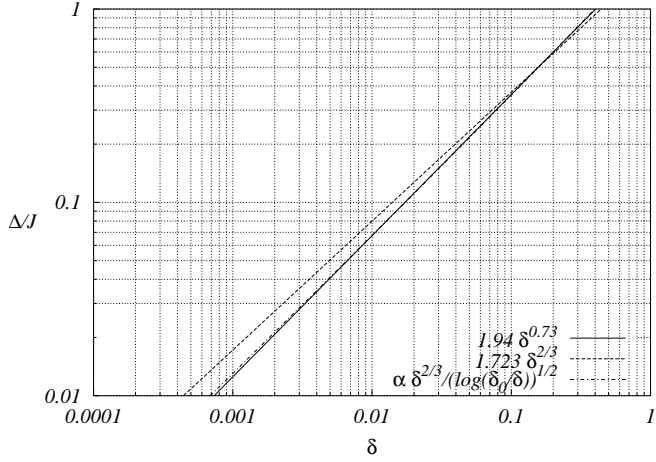}
    \caption{Comparison of the gap obtained in
      Ref.\cite{papenbrock_dimerized} (solid line) with the expressions
      (\ref{eq:gap-predicted}) without logarithmic corrections(dashed
      line)  and the
      expression (\ref{eq:delta-phen}) including logarithmic
      corrections with $y_2(0)=-0.22$
      i.e. $\alpha=4.499$ and $\delta_0=148$ (dash dotted
      line). Equation~(\ref{eq:gap-predicted}) gives a result nearly
      indistinguishable from the result of
      Ref.\cite{papenbrock_dimerized}. }
    \label{fig:compare1}
  \end{center}
\end{figure}

The spin operators can be expressed as a function of $\Pi,\phi$  as:
\begin{eqnarray}
  \label{eq:spin-bosonized}
  \frac{\mathbf{S}_n}{a}&=&(\mathbf{J}_++\mathbf{J}_-)(na) + (-)^{x/a}
  \mathbf{n}(na), \\
  J_r^+(x)&=&(J_r^x+iJ_r^y)(x)=\frac 1{2\pi a} e^{-i\sqrt{2}(\theta-r\phi)(x)}
  \eta_\uparrow \eta_\downarrow,\label{eq:jrplus} \\
  J_r^z&=&\frac 1{2\pi \sqrt{2}} \left[ r\Pi - \partial_x \phi\right],
  \label{eq:jrz}\\
  n^+(x)&=& (n^x+in^y)(x)=\frac{\lambda}{\pi a} e^{-i\sqrt{2}\theta(x)}
  \eta_\uparrow \eta_\downarrow,\label{eq:nplus} \\
  n^z(x)&=&\frac{\lambda}{\pi a} \sin \sqrt{2} \phi_\sigma(x)\label{eq:nz},
\end{eqnarray}
\noindent where $a$ is a lattice spacing, $\eta_{\uparrow/\downarrow}$
represent Majorana fermion operators that can be omitted in some cases
(see e. g. Ref. \cite{schulz_moriond} for a discussion of this
point), and $\theta$ is
defined by $\theta(x)=\pi \int_{-\infty}^x dx' \Pi(x')$.
The constant $\lambda$ is a non-universal parameter that depends on
the lattice model being considered. Recently, this parameter has been
determined in the case of the isotropic Heisenberg spin-1/2
chain\cite{affleck_ampl_xxz,lukyanov_ampl_xxz}, and it was found that:
\begin{eqnarray}
  \label{eq:lambda}
  \lambda=\left(\frac \pi 2 \right)^{1/4}. 
\end{eqnarray}
In order to determine the bosonized Hamiltonian of the dimerized
spin-1/2 chain (\ref{eq:dimerized-spinchain}), all we need is a
bosonized expression of the dimerization operator 
$\sum_n (-)^n \mathbf{S}_n\cdot
\mathbf{S}_{n+1}$.  Using (\ref{eq:spin-bosonized}),
the corresponding expression is easily extracted from:  
\begin{eqnarray}
  \label{eq:current-dimer}
  \frac 1 a \mathbf{S}_{n}\cdot\mathbf{S}_{n+1}&=&(\text{uniform})
  +\nonumber \\ &&
  (-)^n a \left[ - (\mathbf{J}_++\mathbf{J}_-)(na)\cdot \mathbf{n}((n+1)a)
  \right. \nonumber \\ && \left.+\mathbf{n}(na) \cdot (\mathbf{J}_++\mathbf{J}_-)((n+1)a)\right]
\end{eqnarray}
The  bosonized expression of the dimerization operator, is thus
obtained from the
the short distance expansion of $\mathbf{J}_{R,L}$ and $\mathbf{n}$.
 Using Eqs. (\ref{eq:jrplus}) and (\ref{eq:nplus}) with Glauber identities, 
one finds the following expressions:
\begin{subequations}
  \label{eq:sdes-xy}
\begin{eqnarray}
  n^\pm(x)(J_++J_-)^\mp(x+a)&=&\frac{\lambda}{(\pi a)^2} \cos \sqrt{2}
  \phi(x) + \ldots \\ 
  (J_++J_-)^\pm(x)n^\mp(x+a)&=&-\frac{\lambda}{(\pi a)^2} \cos \sqrt{2}
  \phi(x)  + \ldots 
\end{eqnarray} 
\end{subequations} 
The change of sign is a consequence of the application of the Glauber
identity taking into account the commutation relation
$[\phi(x),\theta(x')]=i\pi Y(x'-x)$, $Y$ being the Heaviside step
function. Finally,$n^z(x+a)(J_+^z+J_-^z)(x)$ and $n^z(x)(J_+^z+J_-^z)(x+a)$ 
 are respectively obtained from Eqs.~(\ref{eq:jrz}) and~(\ref{eq:nz})  the two following short distance
expansion: 
\begin{subequations}
  \label{eq:sdes-z}
\begin{eqnarray}
  -\frac 1{2\pi \sqrt{2}} \partial_x \phi (x+a) \sin \sqrt{2} \phi(x)
  &=&\frac 1 {2\pi a} \cos \sqrt{2} \phi(x) +\ldots \\ 
  -\frac 1{2\pi \sqrt{2}} \partial_x \phi (x) \sin \sqrt{2} \phi(x+a)
  &=&-\frac 1 {2\pi a} \cos \sqrt{2} \phi(x) +\ldots  
\end{eqnarray}
\end{subequations}
\noindent which can be derived by normal ordering the product
of the two operators\cite{eggert_openchains,orignac98_spinladder}. A
sketch of the derivation is given in the appendix. It is easily seen that
Eqs.~(\ref{eq:sdes-xy})--~(\ref{eq:sdes-z}) are compatible with spin
rotational invariance.  
Combining the expressions (\ref{eq:sdes-xy}) and (\ref{eq:sdes-z}) in
(\ref{eq:current-dimer}), and using the value of $\lambda$ in Eq.~(\ref{eq:lambda}) we finally obtain that:
\begin{eqnarray}
  \label{eq:dimerization-bosonized}
  \frac 1 a \mathbf{S}_{n}\cdot \mathbf{S}_{n+1}= \text{uniform} +
  (-)^n \frac{3}{\pi^2 a}\left(\frac \pi 2 \right)^{1/4} \cos \sqrt{2} \phi
\end{eqnarray}
Therefore, the continuum Hamiltonian describing the dimerized spin 1/2
chain at low energy reads:
\begin{eqnarray}
  \label{eq:dimer-ham-boso}
  H&=&\int \frac{dx}{2\pi} \left[ uK (\pi \Pi)^2 + \frac u K (\partial_x
    \phi)^2\right] -\frac{2g_1}{(2\pi a)^2} \cos \sqrt{2}\phi\nonumber
  \\ 
  && -\frac{2g_2}{(2\pi a)^2} \int dx \cos \sqrt{8} \phi. 
\end{eqnarray}
Note that in (\ref{eq:dimer-ham-boso}), the sign of $g_1$ does not
matter as it can always been rendered positive by the shift $\phi \to
\phi+\pi/\sqrt{2}$. In (\ref{eq:dimer-ham-boso}), we have:
\begin{eqnarray}
\label{eq:g1-delta}
  g_1=6 J \left(\frac \pi 2\right)^{1/4} \delta a. 
\end{eqnarray}
As we noted before, $g_2$ is a marginally
irrelevant field which flows to $0$ if $g_1=0$.
Let us assume for a moment that we can neglect completely
the presence of this marginally irrelevant operator and take
$K=1,g_2=0$ in (\ref{eq:dimer-ham-boso}). Then, the Hamiltonian
(\ref{eq:dimer-ham-boso}) becomes the sine-Gordon model. 
 This model is integrable, and the expression of the
gap can be found in \cite{zamolodchikov95_gap_sinegordon}, or in
\cite{lukyanov_sinegordon_correlations} 
Eq. (12). In the notations of \cite{lukyanov_sinegordon_correlations},
$\beta^2=K/4=1/4$, and $\mu=3/\pi^3(\pi/2)^{1/4}\delta$ (where we have
used the fact that the velocity $u=\frac \pi 2 J a$). The
dimensionless gap $M$ is then given by:
\begin{eqnarray}
  \label{eq:gap-LZ}
  M=\frac 2 {\sqrt{\pi}} \frac {\Gamma(1/6)}{\Gamma(2/3)}
  \left[\frac{\Gamma(3/4)}{\Gamma(1/4)} \frac 3 {\pi^2} \left(\frac
      \pi 2 \right)^{1/4} \delta\right]^{2/3}\simeq 1.097 \delta^{2/3},
\end{eqnarray}
and the energy gap is given by $\Delta=\frac u a M$ i.e. 
\begin{eqnarray}
  \label{eq:gap-predicted}
  \frac{\Delta}{J}=\frac{\pi} 2 M \simeq 1.723 \delta^{2/3}. 
\end{eqnarray}
We note that the formula (\ref{eq:gap-LZ}) has already been applied
to calculate the gap of the dimerized spin 1/2 chain in
\cite{essler97_ff_sg}, but the value of $\lambda$,
Eq. (\ref{eq:lambda}) was not known. The formula
(\ref{eq:gap-predicted}) is  in reasonable agreement with the result quoted
in \cite{uhrig99_spinons} who reported that
$\Delta/J=1.5\delta^{0.65}$ as two expression differ at most by 6\%
for $0.01\le \delta \le 0.1$.  
Comparing our  expression (\ref{eq:gap-predicted}) to the one of
Ref.\cite{papenbrock_dimerized}, $\Delta/J=1.94\delta^{0.73}$, we find
that they are in agreement within a 10\% relative error when $\delta\ge 0.03$
as represented on  Fig.~\ref{fig:compare1}.  For lower
values of $\delta$, the two results deviate sensibly. As we shall see,
this is the result of the logarithmic corrections.  

In \cite{lukyanov_sinegordon_correlations}, the expression of the
ground state energy was also given in dimensionless units in
Eq. (14). Using this expression, we obtain for the ground state
energy:
\begin{equation}
  \label{eq:energy-LZ}
  \frac{E_0}{J}=-\frac \pi 2 J \frac{M^2}{4} \tan \frac \pi 6 \simeq
  -0.2728 \delta^{4/3}
\end{equation}

\begin{figure}[htbp]
  \begin{center}
    \includegraphics[width=9cm]{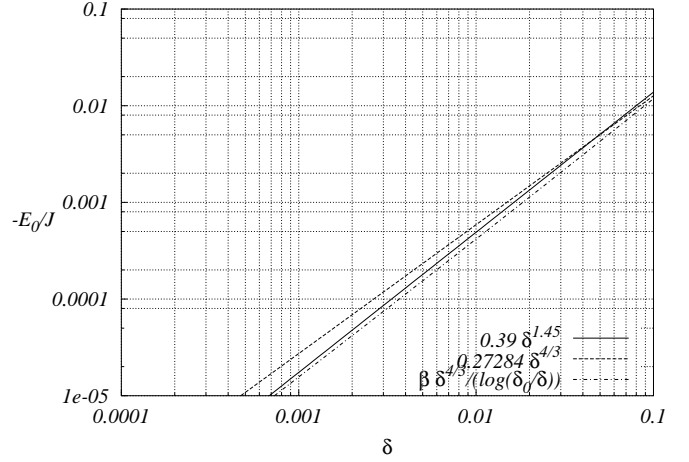}
    \caption{Comparison of the expression of ground state energy of
      Ref. \cite{papenbrock_dimerized} with the expression derived in
      this paper without logarithmic corrections~(\ref{eq:energy-LZ})
      (dashed line) and with logarithmic
      corrections~(\ref{eq:e0-phen}) for $y_2(0)=-0.22$
      i.e. $\beta=1.86$ and $\delta_0=148$ (dash-dotted line). 
      Including logarithmic
      corrections leads to a better agreement for small
      dimerization. However, deviations are still significant in
      contrast with the case of the gap.} 
    \label{fig:compare2}
  \end{center}
\end{figure}

This expression is compared to the one quoted in
\cite{papenbrock_dimerized}, $E_0/J=-0.39 \delta^{1.45}$ on Fig.~\ref{fig:compare2}. The two
energy formulas are in better agreement than the gap formulas at low
dimerization. 
Till now, we have totally neglected the presence of the marginally
irrelevant operator $\cos \sqrt{8}\phi$. As we shall now see, the
 corrections to
scaling\cite{black_equ,affleck_log_corr} induced by this operator 
in the gap formula, are
responsible for the discrepancies between the numerical and the
analytical results. The renormalization group equations associated
with the Hamiltonian (\ref{eq:dimer-ham-boso})
read\cite{kadanoff_ashkin}: 
\begin{eqnarray}
  \label{eq:RGE-dimer}
  \frac d {dl}\left(\frac 1 K\right)&=&\frac 1 8 y_1^2 +\frac 1 2 y_2^2,
  \\ 
  \frac{dy_1}{dl}&=&\left(2-\frac K 2 +y_2\right) y_1, \\
  \frac{dy_2}{dl}&=&(2-2K) y_2 + \frac{y_1^2}{4}, 
\end{eqnarray}
\noindent where we have introduced $y_i=g_i/(\pi u)$.  For $y_1=0$, the
$SU(2)$ symmetric flow is recovered for $K=1-y_2/2$. Then, the
equations (\ref{eq:RGE-dimer}) reduce to the single
Kosterlitz-Thouless\cite{kosterlitz_renormalisation_xy}
$dy_2/dl=y_2^2$. We see that for $y_2<0$, this equation flows to the
fixed point $y_2^*=0$, with the following flow equation:
\begin{eqnarray}
  \label{eq:marginal-scaling}
  y_2(l)=\frac{y_2(0)}{1-y_2(0)l} 
\end{eqnarray}
Let us now assume\cite{kadanoff_ashkin} 
that we have turned on a very small $y_1$. Using the
initial conditions with $SU(2)$ symmetry, we can easily show that the
RG equations reduce to:
\begin{eqnarray}
  \label{eq:RG-y2}
  \frac{dy_2}{dl}&=&y_2^2+\frac 1 4 y_1^2, \\
  \label{eq:RG-y1}
  \frac{dy_1}{dl}&=&\left(\frac 3 2 +\frac 3 4 y_2\right) y_1. 
\end{eqnarray}
If we assume that $y_1(0)\ll y_2(0)$, we can assume that in
(\ref{eq:RG-y2}), we can take $y_1=0$, so that the flow of $y_2$ is
given by (\ref{eq:marginal-scaling}). Then, the equation
(\ref{eq:RG-y1}) is trivially integrated,  leading to:
\begin{eqnarray}
  \label{eq:y1-explicit}
  y_1(l)=y_1(0) \frac{e^{\frac 3 2 l}}{(1+|y_2(0)| l)^{3/4}}
\end{eqnarray}
This equation should break down for a scale $l_0$ such that
$y_1(l_0)\sim y_2(l_0)$. One has:
\begin{eqnarray}
  \label{eq:l0}
  e^{l_0}  {(1+y_2(0)l_0)^{1/6}} =\frac{|y_2(0)|^{2/3}}{y_1(0)^{2/3}}.
\end{eqnarray}
\noindent For $l>l_0$ 
the contribution of $y_2$ to the renormalization of $y_1$ being
negligible, $y_1(l)=e^{3/2(l-l_0)}y_1(l_0)$. The scale
$l^*$ at which $y_1(l)\sim 1$ is thus given by:
\begin{eqnarray}
  \label{eq:lstar-l0}
  e^{-l^*}=e^{-l_0}\frac{(1+|y_2(0)|
    l_0)^{2/3}}{|y_2(0)|^{2/3}}=\frac{|y_1(0)|^{2/3}}{(1+|y_2(0)|l_0)^{1/2}}, 
\end{eqnarray}
 An
approximate form of $l_0$ can be obtained by iterating (\ref{eq:l0})
leading to:
\begin{eqnarray}
  \label{eq:lstar-approx}
  e^{-l^*}\simeq \frac{|y_1(0)|^{2/3}}{\left(1 + \frac 2 3
      |y_2(0)|\ln\left|\frac{y_2(0)} {y_1(0)}
      \right| \right)^{1/2}} 
\end{eqnarray}

Since the scaling of the gap is $\Delta \sim e^{-l^*}$ and the scaling
of the ground state energy is  $E_0\sim e^{-2l^*}$, for $y_1\to 0$,
these formulas are in agreement with the scaling 
predicted in \cite{affleck_log_corr}.  

We now make an important assumption. 
We assume that in
the formulas (\ref{eq:gap-predicted})~(\ref{eq:energy-LZ}), we can
replace $M$ with ${\cal C}e^{-l^*}$, $l^*$ being given by
(\ref{eq:lstar-approx}), 
with ${\cal C}$ being chosen in such a way that for $y_2=0$, the
resulting $M$ agrees with (\ref{eq:gap-LZ}). This is a uncontrolled
approximation as the model (\ref{eq:dimer-ham-boso}) is
non-integrable but it is partially justified by the fact that the energy
and the correlation length evolve continuously as a function of the
parameter $y_2$ in the vicinity of the integrable point. 
With this assumption, and noting that the definition of $y_1$ and
Eq. (\ref{eq:g1-delta}) 
imply $y_1=1.3612 \delta$,  we obtain that ${\cal C}=0.8932$. 

We are lead to the following expressions of the gap:
\begin{equation}
  \label{eq:delta-phen}
  \frac{\Delta}{J}=\frac{1.723 \delta^{2/3}}{\left(1+ \frac 2 3 |y_2(0)| \ln
    \left|\frac{y_2(0)}{1.3612 \delta}\right|\right)^{1/2}}, 
\end{equation}
\noindent and of the ground state energy difference:
\begin{equation}
  \label{eq:e0-phen}
  -\frac{E_0}{J}=\frac{0.2728 \delta^{4/3}}{1+ \frac 2 3 |y_2(0)| \ln
    \left|\frac{y_2(0)}{1.3612 \delta}\right|}
\end{equation}
\noindent These equations are trivially reduced to the Eqs. (7) and (8)
in \cite{papenbrock_dimerized}, by expressing
$\alpha,\alpha_{gap},\delta_0$  as a function of $y_2(0)$. 
A good fit to the power-law expression of the gap quoted  in
\cite{papenbrock_dimerized} is obtained using Eq.~(\ref{eq:delta-phen}) with
$y_2(0)=-0.22$ (see Fig.~\ref{fig:compare1}). 
The corresponding fit for the ground state energy using
Eq.~(\ref{eq:e0-phen}) with the same value of $y_2(0)=-0.22$ is
better than the fit obtained without logarithmic corrections 
(see Fig.~\ref{fig:compare2}), however
it is not as good as the fit obtained for the gap, especially for
$\delta>0.01$. This could be due to the non-singular part of the
free-energy which is not taken into account in the Renormalization
Group calculation.   The value of $y_2(0)=-0.22$  compares
reasonably well with the one quoted in
\cite{affleck_log_corr,schulz_q1daf} $y_0=-0.25$.     

To summarize, we have shown that the results of
Ref.\cite{papenbrock_dimerized} could be recovered from a bosonization
approach including the appropriate operator renormalizations, and
using exact results for the sine-Gordon model combined with a
one-loop RG. The amplitude of the
marginally relevant operator was found to be in reasonable agreement
with an independent estimate coming from logarithmic corrections to
the dependence of the gaps in a spin-1/2 chain. Given the relatively
large value of the coupling constant, we are at the limit of
applicability of the one-loop RG. Better agreement might be obtained
by going beyond the one-loop approximation\cite{amit_xy}. The present
approach does not depend crucially on integrability as it is also
possible to determine the parameter $\lambda$ in
(\ref{eq:spin-bosonized}) for a non-integrable model via numerical
calculations\cite{hikihara03_amplitude_xxz,hikihara_xxz}.  

\begin{acknowledgement}
The author thanks R. Chitra, R. Citro, T. Giamarchi and P. Lecheminant
for discussions.
\end{acknowledgement}

\appendix 
\section{derivation of the short distance expansion}
The short distance expansion (\ref{eq:sdes-z}) has been derived in
\cite{eggert_openchains,orignac98_spinladder} using a fermionic
representation. In this appendix, we give an alternative derivation
using the bosonic representation. 
We start from the equation:
\begin{eqnarray}
  :\partial_x \phi(x'): :V(\phi(x)):=:\partial_x \phi(x')V(\phi(x)): +
  :\frac{dV}{d\phi}(\phi(x)): \partial_x \langle
  \phi(x')\phi(x)-\phi^2\rangle, 
\end{eqnarray}
\noindent which is easily obtained by expanding $V(\phi)$ as a power
series and applying Wick's theorem\cite{bogolyubov_ITQF}. 
In the case of the massless free
boson, we have:
\begin{eqnarray}
   \langle
  \phi(x')\phi(x)-\phi^2\rangle= \frac 1 {2} \ln \left|\frac{x-x'}a\right|,
\end{eqnarray}
\noindent which leads to the expansion:
\begin{eqnarray}
  :\partial_x \phi(x'): :V(\phi(x)):=:\partial_x \phi(x')V(\phi(x)): + \frac{-1}{2(x'-x)}
  :\frac{dV}{d\phi}(\phi(x)): 
\end{eqnarray}
Applying this formula in the case of $V(\phi)=\sin \sqrt{2}\phi$ leads
to Eqs.~(\ref{eq:sdes-z}). 
  

\begin{thebibliography}{10}

\bibitem{bethe_xxx}
H.~A. Bethe, Z. Phys. {\bf 71},  205  (1931).

\bibitem{luther_spin1/2}
A. Luther and I. Peschel, Phys. Rev. B {\bf 12},  3908  (1975).

\bibitem{haldane_xxzchain}
F.~D.~M. Haldane, Phys. Rev. Lett. {\bf 45},  1358  (1980).

\bibitem{nijs_equivalence}
M.~P.~M. den Nijs, Phys. Rev. B {\bf 23},  6111  (1981).

\bibitem{affleck_log_corr}
I. Affleck, D. Gepner, T. Ziman, and H.~J. Schulz, J. Phys. A {\bf 22},  511
  (1989).

\bibitem{giamarchi_logs}
T. Giamarchi and H.~J. Schulz, Phys. Rev. B {\bf 39},  4620  (1989).

\bibitem{barzykin99_logs}
V. Barzykin and I. Affleck, J. Phys. A {\bf 32},  867  (1999).

\bibitem{barzykin01_nmr_logs}
V. Barzykin, Phys. Rev. B {\bf 63},  140412  (2001).

\bibitem{brunel99_edges_logs}
V. Brunel, M. Bocquet, and T. Jolicoeur, Phys. Rev. Lett. {\bf 83},  2821
  (1999).

\bibitem{eggert_susceptibility}
S. Eggert, I. Affleck, and M. Takahashi, Phys. Rev. Lett. {\bf 73},  332
  (1994).

\bibitem{starykh97_logs}
O.~A. Starykh, R.~R.~P. Singh, and A.~W. Sandvik, Phys. Rev. Lett. {\bf 78},
  539  (1997).

\bibitem{nauenberg80_potts}
M. Nauenberg and D.~J. Scalapino, Phys. Rev. Lett. {\bf 44},  837  (1980).

\bibitem{kadanoff_ashkin}
L.~P. Kadanoff, Phys. Rev. B {\bf 22},  1405  (1980).

\bibitem{black_equ}
J.~L. Black and V.~J. Emery, Phys. Rev. B {\bf 23},  429  (1981).

\bibitem{kohmoto81_ashkin}
M. Kohmoto, M. den Nijs, and L.~P. Kadanoff, Phys. Rev. B {\bf 24},  5529
  (1981).

\bibitem{cross_spinpeierls}
M.~C. Cross and D.~S. Fisher, Phys. Rev. B {\bf 19},  402  (1979).

\bibitem{haldane_dimerized}
F.~D.~M. Haldane, Phys. Rev. B {\bf 25},  4925  (1982).

\bibitem{spronken86_dimerized}
G. Spronken, B. Fourcade, and Y. {L\'epine}, Phys. Rev. B {\bf 33},  1886
  (1986).

\bibitem{kung86_dimerized}
D. Kung, Phys. Rev. B {\bf 34},  6315  (1986).

\bibitem{barnes_dimerized}
T. Barnes, J. Riera, and D.~A. Tennant, Phys. Rev. B {\bf 59},  11384  (1999).

\bibitem{uhrig99_spinons}
G. Uhrig, F. Schönfeld, M. Laukamp, and E. Dagotto, Eur. Phys. J. B {\bf 7},
  67  (1999).

\bibitem{singh99_dimerized}
R.~R.~P. Singh and Z. Weihong, Phys. Rev. B {\bf 59},  9911  (1999).

\bibitem{chitra95_dmrg_frustrated}
R. Chitra {\it et~al.}, Phys. Rev. B {\bf 52},  6581  (1995).

\bibitem{papenbrock_dimerized}
T. Papenbrock {\it et~al.}, Phys. Rev. B {\bf 68},  024416  (2003).

\bibitem{affleck_ampl_xxz}
I. Affleck, J. Phys. A {\bf 31},  4573  (1998).

\bibitem{lukyanov_ampl_xxz}
S. Lukyanov, Phys. Rev. B {\bf 59},  11163  (1999).

\bibitem{hikihara03_amplitude_xxz}
T. Hikihara and A. Furusaki, cond-mat/0310391 (unpublished).

\bibitem{schulz_moriond}
H.~J. Schulz,  in {\em Correlated fermions and transport in mesoscopic
  systems}, edited by T. Martin, G. Montambaux, and J. {Tran Thanh Van}
  (Editions fronti\`eres, Gif sur Yvette, France, 1996), p.\ 81.

\bibitem{eggert_openchains}
S. Eggert and I. Affleck, Phys. Rev. B {\bf 46},  10866  (1992).

\bibitem{orignac98_spinladder}
E. Orignac and T. Giamarchi, Phys. Rev. B {\bf 57},  5812  (1998).

\bibitem{zamolodchikov95_gap_sinegordon}
Al.~B.  Zamolodchikov, Int. J. Mod. Phys. A {\bf 10},  1125  (1995).

\bibitem{lukyanov_sinegordon_correlations}
S. Lukyanov and A.~B. Zamolodchikov, Nucl. Phys. B {\bf 493},  571  (1997).

\bibitem{essler97_ff_sg}
F.~H.~L. Essler, A.~M. Tsvelik, and G. Delfino, Phys. Rev. B {\bf 56},  11001
  (1997).

\bibitem{kosterlitz_renormalisation_xy}
J.~M. Kosterlitz, J. Phys. C {\bf 7},  1046  (1974).

\bibitem{schulz_q1daf}
H.~J. Schulz, Phys. Rev. Lett. {\bf 77},  2790  (1996).

\bibitem{amit_xy}
D.~J. Amit, Y.~Y. Goldschmidt, and G. Grinstein, J. Phys. A {\bf 13},  585
  (1980).

\bibitem{hikihara_xxz}
T. Hikihara and A. Furusaki, Phys. Rev. B {\bf 58},  R853  (1998).

\bibitem{bogolyubov_ITQF}
N.~N. Bogolyubov and D.~V. Shirkov, {\em {Introduction to the theory of
  quantized fields}} (Wiley Interscience, New York, 1957).

\end{thebibliography}

\end{document}